\documentclass[aps]{revtex4}
\usepackage{epsf,graphicx}
\usepackage{amsmath}
\usepackage{amssymb}
\newcommand{\bb}{\begin{eqnarray}}
\newcommand{\ee}{\end{eqnarray}}
\usepackage{subfig}
\usepackage{graphicx}
\begin{document}
\title{\bf Strong gravitational lensing across dilaton anti-de Sitter black hole }
\author{Tanwi Ghosh\footnote{E-mail: tanwi.ghosh@yahoo.co.in} and Soumitra SenGupta\footnote{E-mail: tpssg@iacs.res.in}}
\affiliation{Department of Theoretical Physics , Indian Association for the
Cultivation of Science,\\
Jadavpur, Calcutta - 700 032, India}
\vskip 5cm
 
\begin{abstract}
In this work we investigate  gravitational lensing effect in strong field region 
around a dilaton black holes in an anti de Sitter ( ADS ) space. We also analyse the dependence of the radius of the photon sphere and 
deflection angle on dilaton coupling and cosmological constant in this black hole space time. Finally the
values of  minimum impact parameter , 
the separation between the first and the other images as well as  
the ratio between the flux of the first image and the flux coming from all the other images are determined to characterize 
some possible distinct signatures of such black holes.
\end{abstract}
\maketitle

{{\Large {\bf 1. Introduction :}}}\\

One of the key signature of astrophysical objects like black holes is the gravitational lensing effect.
Gravitational lensing is the deflection of electromagnetic radiation in a gravitational field. 
Intense research on the phenomena of gravitational lensing is being carried out in recent times. 
The results of these studies also have their relevance in the detection of extrasolar planets and  in compact dark matter 
to estimate the value of the cosmological constant.
When the gravitational field is strong,i.e light rays pass very close to a black hole, observer can detect infinite 
series of images formed very close to black hole. 
An incoming photon with an impact parameter, deviates as it approaches a minimum distance towards the black hole.
In such a case deflection angle increases with decrease of the impact parameter.The photon will traverse a complete
loop across the black hole when the deflection angle exceeds $2\pi$. For further lowering of the value of the impact
parameter,it will execute several windings around the black hole. It is shown \cite{vir1} that the deflection
angle diverges as light rays approach towards the photon sphere.
This phenomena is known as gravitational lensing in strong deflection region. Very long baseline interferometry may be able to 
detect relativistic images and the effects of strong fields within these objects\cite{con}. Recently various 
works have been done on gravitational lensing in both 
strong and weak field region\cite{li,ref,bour,sch,surdej,wam,dar,bekenstein,fri,ks,majum,rom,bh,john,morganson,song,
koop, faure,bis}.
Strong gravitational lensing can provide informations about the nature of spacetime around 
various kinds of black holes. Our aim in this work is to study the strong field  gravitational lensing phenomena  near a dilaton coupled 
black hole in presence of cosmological constant \cite{gao}. Here 
we have explored the role of cosmological constant on deflection angle in the limit of  small dilaton 
coupling and have obtained the values of minimum impact parameter ,radius of the 
photon sphere , the separation between the first and the other images as well as  the ratio 
between the flux of the first image and the flux coming from all the other images.
One of the key motivation of the present study is that in presence of dilaton field, black holes have distinct 
features which are different from other black holes. Moreover such black holes are interesting to study because 
the fate of dark energy dominated universe may be influenced when dynamical modulus or 
dilaton fields coupled to string curvature are taken into account \cite{sami}and it is suggested that our accelerating 
universe may be dominated by the energy density of dilaton field\cite{gas}.\\
The gravitational action for dilaton black holes in anti-de Sitter spacetime can be described as,\cite{gao}:\\

\begin{eqnarray}
S=\int d^4x \sqrt{-g}[R-2\partial_{\mu}\varphi\partial^{\mu}\varphi-e^{-2a\varphi}F^2
-V(\varphi)]
\end{eqnarray}
where R is the Ricci scalar, 
$V(\varphi)=\frac{-2a}{3(1+a^2)^2}[a^2(3a^2-1)e^{\frac{-2\varphi}{a}}+(3-a^2)e^{2a\varphi}+8a^2e^{a\varphi-\frac{\varphi}{a}}]$ is the 
potential term in presence of 
cosmological constant $\lambda$ for dilaton anti de Sitter black hole ($\lambda<0$), $\varphi$ is the 
dilaton field and '$a$' represents coupling parameter of the dilaton with Maxwell field $F_{\mu\nu}$.\\

{{\Large {\bf 2.Strong Field Lensing By Dilaton Anti-de Sitter Black hole :}}}\\

Considering dilaton scalar filed $\varphi$ coupled to electromagnetic field of 
charge $Q$ with arbitrary coupling parameter '$a$' in presence of a
Liouville type potential containing the cosmological constant $\lambda$, Gao and Zhang obtained 
the dilaton anti-de Sitter black hole solutions.
To study strong field lensing occurring around such dilaton anti-de Sitter black holes, we consider the following spacetime\cite{gao}
\begin{eqnarray}
ds^2=-A(r)dt^2+B(r)dr^2+C(r)d\Omega^2
\end{eqnarray}
where $A(r)=[(1-\frac{r_+}{r})(1-\frac{r_-}{r})
^{\frac{(1-a^2)}{(1+a^2)}}-\frac{\lambda}{3}r^2((1-\frac{r_-}{r})
^{\frac{2a^2}{(1+a^2)}}]$,$B(r)=[(1-\frac{r_+}{r})(1-\frac{r_-}{r})
^{\frac{(1-a^2)}{(1+a^2)}}-\frac{\lambda}{3}r^2((1-\frac{r_-}{r})
^{\frac{2a^2}{(1+a^2)}}]^{-1}$, 
$C(r)=r^2(1-\frac{r_-}{r})^{\frac{2a^2}{(1+a^2)}}$,
$r_-r_+=(1+a^2)Q^2e^{2a\varphi_0}$, 
the expression of dilaton field with asymptotic value $\varphi_0$ is
$e^{2a\varphi}=e^{2a\varphi_0}(1-\frac{r_-}{r})^{\frac{2a}{(a^2+1)}}$
and the only non-vanishing component of $F_{\mu\nu}$ is
$F_{01}=\frac{Q^2 e^{2a\varphi}}{C}$

In order to explain strong field lensing effect around black holes,we need to obtain the expressions of radius of the photon sphere
and also the deflection angle. Using these two variables one can then determine other observables such as separation between the 
first and other images, the ratio between the flux of the first image and the flux coming from all the other images.\\

{{\Large {\bf Radius of the photon sphere:}}}\\

Two different methods has been used \cite{vir1,vir2} and obtained the following condition to determine the radius of 
the photon sphere as,

\begin{eqnarray}
\frac{C'}{C}=\frac{A'}{A}
\end{eqnarray}
After some algebraic calculation and taking leading order terms in $\frac{1}{r}$,one can obtain the radius of the photon 
sphere in the strong field region as, 
\begin{eqnarray}
r_{ps}=\frac{k_2+\sqrt{k_2^2-4k_1k_3}}
{2k_1}
\end{eqnarray}
where $k_1=[2(1+a^2)-\frac{4a^2 r_-^2\lambda}{3(1+a^2)}+\frac{4\lambda}{3}r_-^2
a^2-2a^2r_-^2\frac{\lambda}{3}-\frac{(a^2-1)}{(1+a^2)}2a^2 r_-^2
\frac{\lambda}{3}]$,\\
$k_2=[2r_+(1+a^2)+2r_-(1-a^2)+2r_-+(1+a^2)r_++(1-a^2)r_--
\frac{2a^2\lambda}{3}r_-^3\frac{(a^2-1)}{(1+a^2)}]$,\\
$k_3=[2(1-a^2)r_-r_++2r_-r_++2r_-^2\frac{(1-a^2)}{((1+a^2)}+(1+a^2)r_-r_+
+2(1-a^2)r_-r_++(1-a^2)r_-^2-\frac{2a^2(1-a^2)}{(1+a^2)}r_-^2]$.\\
Equation (4) exhibits the dependence of $r_{ps}$ on cosmological constant $\lambda$ as well as dilaton coupling '$a$'.Variation of $r_{ps}$
with dilaton coupling '$a$' is shown in the following figure.

\begin{figure}[h] 
\includegraphics[width=3.330in,height=2.20in]{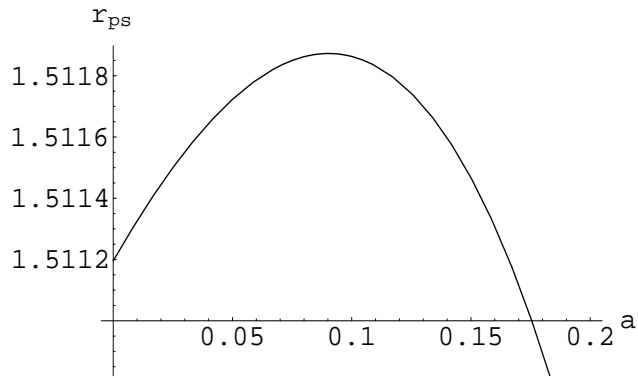}
\caption{Graph of the dilaton coupling '$a$' versus the radius of the photon sphere  $r_{ps}$} \label{radphot}
for Q=0.3 ,$\varphi_0=1$ and $\lambda=0.1$
\end{figure}
It is evident that the radius of the photon sphere at first increases with increase in the dilaton coupling parameter '$a$',
becomes maximum and then decreases with coupling parameter '$a$'. It has also been found that $r_{ps}$ becomes imaginary
for coupling '$a \sim 0.8$'. Similar situation arises in lensing effect near a Reissner-Nordstrom ( RN ) 
black hole when the charge $Q > \frac{1}{2}$ \cite{vir2}. From equation (4),one can find that the 
dependence of $r_{ps}$ on $\lambda$ is nearly flat.
Setting $r_-=0$,'$a$' = $0$ and $\lambda=0$, in the above expression, we can retrieve  the Schwarzschild value of $r_{ps}$.
We consider the strong field expansion in presence of a photon sphere. A photon, coming from infinity
has an impact parameter $u$ and
when it approaches towards the black hole and reaches the distance of closest approach $r_0$,  it deviates away from the black hole .\\
Utilising the principle of conservation of angular momentum, the impact parameter u can be expressed in terms of 
the distance of closest approach $r_0$\cite{vir3} as\\
\begin{eqnarray}
u=\sqrt{\frac{C_0}{A_0}}
\end{eqnarray}

{{\Large {\bf Deflection Angle:}}}\\

Let us now calculate the deflection angle for the above dilaton-anti de Sitter black hole. 
We have seen from\cite{vir1} that the deflection angle can be increased by decreasing the impact parameter where the
authors have shown the variation of deflection angle w.r.t impact parameter for various values of $\nu$ of JNW
black holes\cite{janis}.The deflection angle becomes enourmously large when the impact parameter approaches towards
the value of the impact parameter of the corresponding photon spehere.
However at some point it exceeds $2\pi$. At this point it will make a complete loop enclosing the black hole. 
Finally for $r_0=r_{ps}$, deflection angle will diverge and the circulating photon can be captured.
To find out analytical behaviour of deflection angle in strong field region \cite{boz}in terms of black hole parameters as well as cosmological constant,let us at first introduce a regularised integral $I_R(r_0)$ for this dilaton-anti de Sitter 
black hole spacetime as,

\begin{eqnarray}
I_R(r_0)
=\int_{0}^{1}[R(z,r_0)f(z,r_0)-R(0,r_{ps})f_0(z,r_0))]\\
=\int_{0}^{1}\frac{2 dz}{r_0\sqrt{d}(1-\frac{r_-(1-z)}{r_0})^{\frac{2a^2}{(a^2+1)}}\sqrt{a_1z+bz^2}}-\int_{0}^{1}\frac{2 dz}{r_0\sqrt{d}(1-\frac{r_-}{r_0})^{\frac{2a^2}{(a^2+1)}}\sqrt{a_1z+bz^2}}
\end{eqnarray}
where
\begin{eqnarray}
d=\frac{(1-\frac{r_+}{r_0})(1-\frac{r_-}{r_0})
^{\frac{(1-a^2)}{(1+a^2)}}-\frac{\lambda}{3}r_0^2((1-\frac{r_-}{r_0})
^{\frac{2a^2}{(1+a^2)}}}{r_0^2(1-\frac{r_-}{r_0})^{\frac{2a^2}{(1+a^2)}}}\\
R(z,r_0)= \frac{2 }{r_0\sqrt{d}(1-\frac{r_-(1-z)}{r_0})^{\frac{2a^2}{(a^2+1)}}}\\
R(0,r_0)= \frac{2 }{r_0\sqrt{d}(1-\frac{r_-}{r_0})^{\frac{2a^2}{(a^2+1)}}}\\
f(z,r_0)\approx f_0(z,r_0)=\frac{1}{\sqrt{a_1z+bz^2}}\\
l=(\frac{2}{d r_0^2}+2\frac{4a^4\lambda r_-^2}{3dr_0^2(a^2+1)^2})\\
b=-\frac{1}{d r_0^2}-\frac{4a^4\lambda r_-^2}{3dr_0^2(a^2+1)^2})\\
c=1-\frac{1}{dr_0^2}+\frac{\lambda}{3d}-\frac{4a^4\lambda r_-^2}{3dr_0^2(a^2+1)^2})\\
a_1=l+2b\beta\\
\beta==\frac{-l+\sqrt{l^2-4bc}}{2b}
\end{eqnarray}
Integrating equation (7) we get
\begin{eqnarray}
I_R(r_0)\nonumber=-\frac{2}{r_0\sqrt{d\alpha_1}}[log[(\frac{1}{p(1-\beta)+q}+\frac{\beta_1}{2\alpha_1})+\sqrt{(\frac{1}{p(1-\beta)+q}+\frac{\beta_1}{2\alpha_1})^2+(b-\frac{\beta_1^2}{4\alpha_1^2})}]-log[(\frac{1}{-p\beta+q}+\\
\nonumber\frac{\beta_1}{2\alpha_1})+\sqrt{(\frac{1}{-p\beta+q}+\frac{\beta_1}{2\alpha_1})^2+(b-\frac{\beta_1^2}{4\alpha_1^2})}]]\\
\nonumber-\frac{2}{r_0\sqrt{bd}(1-\frac{2a^2 r_-}{(\alpha^2+1)r_0})}[log((1-\beta+\frac{a_1}{2b})+\sqrt{(1-\beta+\frac{a_1}{2b})^2+ (\frac{a_1}{2b})^2})\\-log((-\beta+\frac{a_1}{2b})+\sqrt{(-\beta+\frac{a_1}{2b})^2+(\frac{a_1}{2b})^2})]
\end{eqnarray}
where
$p=\frac{2a^2 r_-}{(a^2+1)r_0}$,$q=(1-\frac{2a^2 r_-}{(a^2+1)r_0}+\frac{2a^2 r_-\beta}{(a^2+1)r_0})$,$\alpha_1=(bq^2-(l+2b\beta)pq)=(bq^2-a_1pq)$,$\beta_1=(a_1p-2bq)$ .\\
The deflection angle $\alpha(\theta)$ in the strong field region in terms of impact parameter $u_{ps}$,the angular separation of the image from the lens  $\theta$ and the distance between the lens and the 
observer $D_{0L}$, can be written as\cite{vir3}:\\
\begin{eqnarray}
\alpha(\theta)=-\bar{a}log(\frac{\theta D_{0L}}{u_{ps}}-1)+\bar{b}
\end{eqnarray}
Using the expression of $I_R(r_0)$, at $r_0=r_{ps}$, one can easily find 
\begin{eqnarray}
\bar{a}=\frac{1}{r_{ps}\sqrt{d}(1-p)\sqrt{b}}
\end{eqnarray}
and
\begin{eqnarray}
\bar{b}=-\pi +b_R+\bar{a}log(\frac{2b_{ps}}{A_{ps}})
\end{eqnarray}
where $b_R$ can be expressed through the regularised integral as
\begin{eqnarray}
b_R = I_R(r_{ps})
\end{eqnarray}
Substituting $\bar{a}$,$\bar{b}$,$u_{ps}$,one arrives at the general expression for deflection angle 
for any value of '$a$' in the strong field region.
Considering dilaton coupling parameter '$a$'$<<1$, we have already found the radius of the photon sphere where 
in the leading order the contribution
due to $\lambda$ turned out to be  negligible. 
We then apply it in determining the deflection angle as a function of cosmological constant $\lambda$ 
in strong field region.
The expression of deflection angle in this approximation can be written as,
\begin{eqnarray}
\alpha(\theta)&=&\frac{1}{(1-\frac{2a^2r_-}{(a^2+1)r_{ps}})}log[\frac{2}{dr_{ps}^2n}]-3.14
\nonumber-\frac{1}{(1-\frac{2a^2r_-}{(a^2+1)r_{ps}})}log[\frac{0.003\sqrt{n}}{r_{ps}(1-\frac{r_-}{r_{ps}})^{\frac{a^2}{(1+a^2)}}}]\\
&&-\frac{2}{r_{ps}\sqrt{d\alpha_1}}[log(-\frac{4a^4r_-^2(1-\beta)^2}{r_{ps}^2[1-\frac{4a^4r_-^2
\nonumber(1-\beta)^2}{r_{ps}^2}]}+\frac{2a^2r_-(1-\beta)}{r_{ps}[1-\frac{4a^4r_-^2(1-\beta)^2}{r_{ps}^2}]})\\
&&-log(\frac{2}{(1-\frac{2a^2r_-}{r_{ps}})}-\frac{2}{[1-\frac{4a^4r_-^2(1-\beta)^2}{r_{ps}^2}]})]
-\frac{2}{r_{ps}\sqrt{bd}(1-\frac{2a^2r_-}{r_{ps}})}[log(1-\beta)-log(\sqrt{1+(1-\beta)^2}-1)]
\end{eqnarray}
where $n$ and $\beta$ in this approximation reduce to
\begin{eqnarray}
n=(1-\frac{1}{r_{ps}})(1-\frac{r_-}{r_{ps}})-\frac{\lambda}{3}r_{ps}^2(1-\frac{r_-}{r_{ps}})^2
\end{eqnarray}
and
\begin{eqnarray}
\beta=\frac{\frac{2(1-\frac{r_-}{r_{ps}})^2}{n}-\frac{2(1-\frac{r_-}{r_{ps}})}{\sqrt{n}}}{\frac{2(1-\frac{r_-}{r_{ps}})^2}{n}}
\end{eqnarray}
Variation of $\alpha(\theta)$ with cosmological constant $\lambda$ is shown in Fig.(2) and Fig.(3)for 
different dilaton coupling and charge Q taking $\varphi_0=0$. For n to be real, 
equation (23) implies a bound for the cosmological constant $\lambda$ which depends on the values of the parameters
'$a$' and '$Q$'.
For a fixed value of $\lambda$ the deflection angle increases as 
dilaton coupling '$a$' increases. Comparision of Figures for 
same coupling parameter, say '$a$'=$0.02$, but different values 
of charge Q=0.1 and 0.3 reveals the fact that deflection angle decreases as charge Q increases in the approximation '$a$'$<<1$.\\


\begin{figure}
\centering
\subfloat{\includegraphics[width=0.4\textwidth]{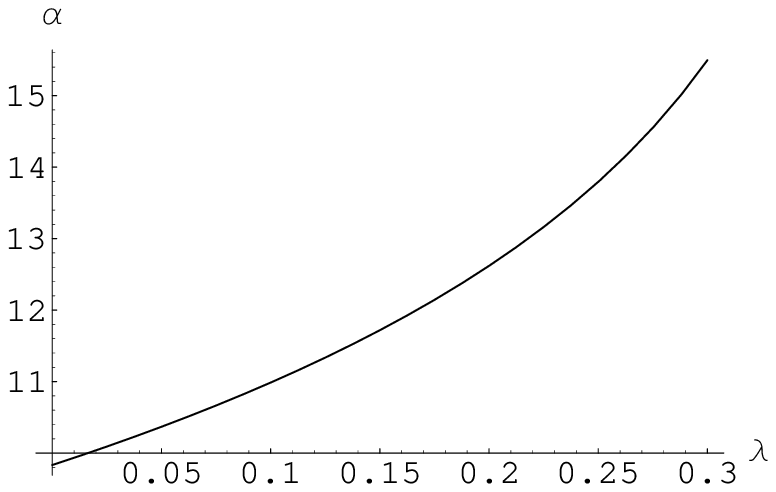}}                
\subfloat{\includegraphics[width=0.4\textwidth]{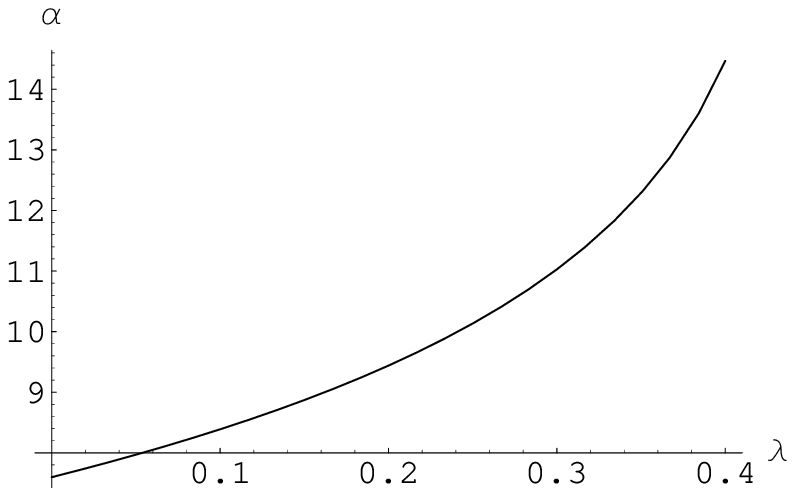}}
\caption{Graphs of cosmological constant $\lambda$ versus deflection angle $\alpha$ for coupling constant 
'$a$'= $0.2$ and charge $Q = 0.1$ and '$a$' = $0.02$ and charge $Q = 0.1$}
\end{figure}



\begin{figure}
\centering
\subfloat{\includegraphics[width=0.4\textwidth]{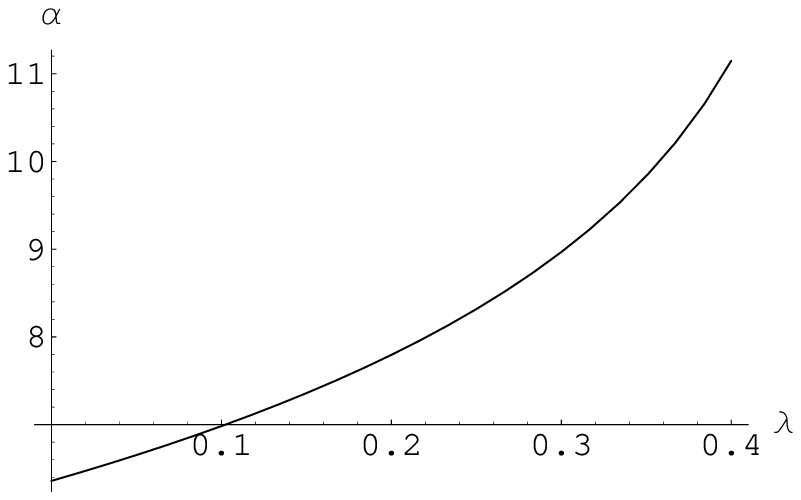}}               
\subfloat{\includegraphics[width=0.4\textwidth]{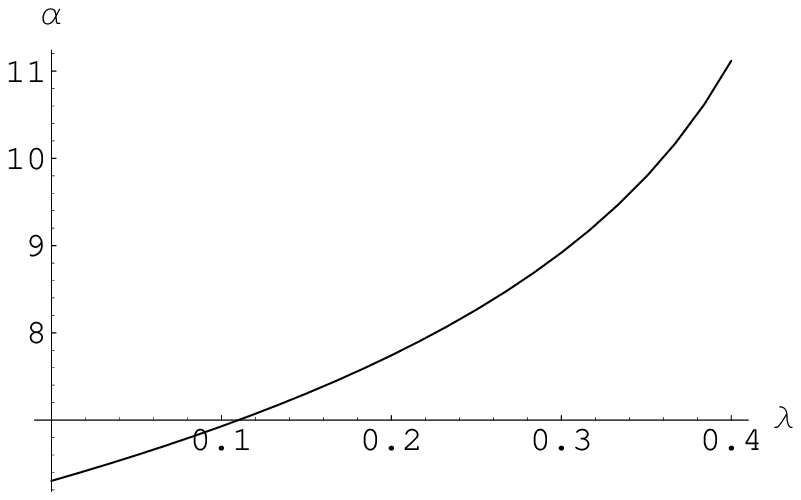}}
\caption{Graphs of cosmological constant $\lambda$ versus deflection angle $\alpha$ for coupling constant 
'$a$' = $0.2$ and charge $Q = 0.3$ and '$a$' = $0.02$ and charge $Q = 0.3$}
\end{figure}

{{\Large {\bf Observables in strong field region:}}}\\

We now estimate the values of minimum impact parameter $u_{ps}$, the radius of the 
photon sphere $r_{ps}$, separation between the first and the other images $s$, the ratio between the flux of the first image and the 
flux coming from all the other images $r$ and the coefficients $\bar{a}$ and $\bar{b}$. In our case we have 
considered the coupling parameter '$a$'$<<1$. Using appropriate formula for $r$ and $s$, 
\begin{eqnarray}
r=exp{(\frac{2\pi}{\bar{a}})}
\end{eqnarray}
\begin{eqnarray}
r_m=2.5log r
\end{eqnarray}
and
\begin{eqnarray}
s=\Theta_{\infty}exp{(\frac{\bar{b}}{\bar{a}}-\frac{2\pi}{\bar{a}})}
\end{eqnarray}
we have obtained the values of observables seting the asymptotic position of a set of images $\Theta_{\infty}=16.87$ (Schwarzschild value), Q=0.3 and
$\lambda=0.2$ as follows.\\

{{\Large {Table: Observables in strong field region taking Q=0.3 and $\lambda=0.2$:}}}\\

\begin{center}
  \begin{tabular}{ l | c | c | c | c | c | c | c | c | c | r | }
    \hline
    coupling
    parameter & $u_{ps}$ & $r_{ps}$ & $\bar{a}$ & $\bar{b}$ & $s$ $\mu$arcsec & $r_m$ (mag) & $\alpha(\theta)$rad \\ \hline
    a=0.2 & 3.3097 & 1.506 & 1.0049 & 0.7534 &  0.06897 & 6.785178 & 7.79\\ \hline
    a=0.02 & 3.3171 & 1.5081 & 1.0000478 & 0.7335 & 0.06583 & 6.818087 & 7.74\\
    \hline
  \end{tabular}
\end{center}
From this Table it is observed that for the charged dilaton black hole in presence of cosmological constant
the values of impact parameter $u_{ps}$ as well as the radius of the photon sphere $r_{ps}$ are larger than the corresponding 
Schwarzschild values. Values of $\bar{a}$,$\bar{b}$, $s$, $r$ and deflection angle $\alpha(\theta)$ are also different. We observe that as 
we decrease the dilaton coupling parameter '$a$' with electromagnetic field ,$u_{ps}$, $r_{ps}$,as well as $r$
will increase while the separation between the images $s$, deflection angle $\alpha(\theta)$, $\bar{a}$
and $\bar{b}$ will decrease.\\
Dependence of n-th image position $\Theta_n$ and their n-th magnification $\mu_n$ on angular source position
$\beta$ is given qualitatively by the following equations,
\begin{eqnarray}
\Theta_n=\Theta_n^0+\frac{u_{ps}(\beta-\Theta_n^0)D_{OS}exp{(\frac{\bar{b}}{\bar{a}}-\frac{2n\pi}{\bar{a}})}}{\bar{a} D_{OL}D_{LS}}
\end{eqnarray}
and
\begin{eqnarray}
\mu_n=exp{(\frac{\bar{b}}{\bar{a}}-\frac{2n\pi}{\bar{a}})}\frac{u_{ps}^2D_{OS}(1+exp{(\frac{\bar{b}}{\bar{a}}-
\frac{2n\pi}{\bar{a}})})}{\bar{a}\beta D_{OL}^2D_{LS}}
\end{eqnarray}
where $\Theta_n^0=\frac{u_{ps}(1+exp{(\frac{\bar{b}}{\bar{a}})})}{D_{OL}}$ for $\alpha{(\theta)}=2n\pi$,$D_{LS}$,$D_{OS}$
and $D_{OL}$ are the lens-source,observer-source and observer-lens distances respectively.For images on the opposite side
of the source,one has to substitute $-\beta$ in equation (28).  Sustituting the values of
$\bar{a}$,$\bar{b}$ and $r_{ps}$ for coupling parameter a=0.2,Q=0.3,$\lambda=0.2$,$\varphi_0=0$ and $D_{OS}=D_{OL}+D_{Ls}
=2D_{OL}$, we obtain the variation of image position (say for primary and secondary images) on the same side(s) and opposite side (o)of the source and their magnification with respect to angular source position as follows,
\begin{eqnarray}
\Theta_{ps}=\frac{3.3232}{D_{OL}}+\frac{0.02693}{D_{OL}}(\beta-\frac{3.3232}{D_{OL}})\\
\Theta_{ss}=\frac{3.30973}{D_{OL}}+\frac{0.000052}{D_{OL}}(\beta-\frac{3.30973}{D_{OL}})\\
\Theta_{po}=\frac{3.3232}{D_{OL}}-\frac{0.02693}{D_{OL}}(\beta+\frac{3.3232}{D_{OL}})\\
\Theta_{so}=\frac{3.30973}{D_{OL}}-\frac{0.000052}{D_{OL}}(\beta+\frac{3.30973}{D_{OL}})\\
\mu_p=\frac{0.0895}{\beta D_{OL}^2}\\
\mu_{s}=\frac{0.0001722}{\beta D_{OL}^2}
\end{eqnarray}
Above equations reveal the fact that the dependence of angular image positions and their magnifications with respect to angular source positions are respectively linear and hyperbolic in nature as was also shown in\cite{vir4,vir5}.
From equations (30) and (31) we notice that the angular position on the same side of the source of primary (ps) and 
secondary(ss) relativistic images increase and decrease respectively with the increment of angular source position 
$\beta$ whereas equations(32) and (33) mention the case of decrease of both primary and secondary angular image
positions on the opposite side of the source as $\beta$ increases.
Equations (34) and (35) imply the decrease of magnifications of the corresponding images as $\beta$ increases. As we decrease the dilaton coupling '$a$', $\Theta_p$ will increase with $\beta$,where as $\mu_p$ will decrease for decrease of '$a$' for a fixed value of $\beta$.\\
Substituting $\beta=0$ in equation (28),i.e,the source,the observer and the lens are in a line ,Einstein's ring 
\cite{vir1,vir3,vir4,vir5}for dilaton coupling $a=0.2$ may be produced as follows,
\begin{eqnarray}
\Theta_{E}=\frac{3.3232}{D_{OL}}-\frac{0.02704}{D_{OL}^2}
\end{eqnarray}
Einstein's ring decreases slowly with a decrease in the value of $a$.
The above qualitative studies of $\Theta_n$ , $\mu_n$ and $\Theta_E$ agree with \cite{vir4,vir5}.\\

 Finally the features of strong gravitational lensing by naked singulaities may be observed from $r_{ps}$ vs
dilaton coupling curve. For the black holes considered in this work, 
SNS(strongly naked singularity) is generated approximately above '$a$'=0.15 and below '$a$'=0.2 where as  WNS(weakly
naked singularity )region lies approxmately within the region  '$a$'=0.15 to  $0.18$.  
Such classifications of weakly and strongly naked singularities was done by Virbhadra and Ellis \cite{vir1}. According to them 
the qualitative features of geodesics are similar for Schwarzschild black hole as well as for WNS.
These result into one Einstein ring and no radial critical curve whereas SNS generates two or nil Einstein ring(s)and also one 
radial critical curve.The authors also mentioned about infinite number of relativistic images
for WNS but no relativistic images for SNS. Moreover Virbhadra and Keeton\cite{vir5} classify WNS and SNS by considering the nature of
time delays of relativistic images. The images for Schwarzschild black hole and WNS have positive time delays but SNS
generates positive, zero as well as negative time delays. SNS always gives rise to negative time delay of the direct image when the 
angular source position is large.\\

{{\Large {\bf 3.Conclusion :}}}\\

In this work we have extensively studied gravitational lensing effect around dilaton anti de Sitter black hole. 
Such kind of Ads black holes are interesting to study because recently quantum effects on black hole geometries
have been incorporated in the context of ADS/CFT correspondence. Moreover five dimensional brane model proposed 
by Randall and Sundrum with infinite warped extra dimension in a Ads bulk has been considered to be equivalent to 
four dimensional Einstein gravity \cite{randall,her,flachi,sddmssg}.
In our present work in the context of strong field lensing around dilaton black holes in presence of cosmological
constant, the method of gravitational lensing effect has been  utilised in 
determining $r_{ps}$,$\alpha(\theta)$,$\bar{a}$,$\bar{b}$,$r$ and $s$ for some selected values of cosmological constant.
From Fig (1)we find  that the 
radius of the photon sphere $r_{ps}$ at first increases with the dilaton coupling '$a$', becomes maximum and then 
again decreases.
We have also explored the variation of deflection angle with cosmological constant $\lambda$ for 
dilaton coupling parameter '$a$'$<<1$. Different plots in Figs (2) and (3) also reveal  that the deflection angle $\alpha_{\theta}$ 
increases with the dilaton coupling parameter '$a$'
when $\lambda$ is kept fixed. We have also explored the variation of the deflection
angle with the black hole charge $Q$.  Taking  '$a$'=0.02, it is shown that $\alpha_{\theta}$ decreases as $Q$ increases.
We have calculated several strong field lensing parameters such as separation between the first and other images,the
ratio between the flux of the first and other images, minimum impact parameter and constant parametres $\bar{a}$ and 
$\bar{b}$ taking $\Theta_{\infty}=16.87$.
The values of $u_{ps}$ ,$\bar{a}$,$\bar{b}$,$r$ and $s$ for such kind of anti-de Sitter black holes are found to be 
distinct from those obtaned for Schwarzschild and dilaton black holes\cite{bh} and therefore can be useful signature to detect such dilaton coupled anti-de Sitter black hole.\\

{{\Large {\bf 4.Acknowledgement :}}}\\

T.G wishes to thank DST (Govt Of India) for financial support.\\


\begin{thebibliography}{99}
\bibitem{vir1}K.S.Virbhadra and G.F.R.Ellis,Phys.Rev.{\bf 65},103004,2002.
\bibitem{con}Constellation-X web page:constellation.gsfc.nasa.gov;maxim web page:maxim.gsfc.nasa.gov;J.S.Ulvestad
astro-ph/9901374.
\bibitem{li}S.Liebes,Jr.,Phys.Rev.{\bf 133},B835,1964.
\bibitem{ref}S.Refsdal,Mon.Not.R.Astron.Soc{\bf 128},295,1964.
\bibitem{bour}R.R.Bourassa and R.Kantowski,Astrophys.J.{\bf 195},13,1975.
\bibitem{sch}P.Schneider,J.Ehlers and E.E.Falco,Gravitational Lenses(Springer-Verlag,Berlin,1992).
\bibitem{surdej}S.Refsdal and J.Surdej,Rep.Prog.Phys{\bf 57},117,1994.
\bibitem{wam}J.Wambsganss,Living Rev.Relativity{\bf 1},12,1998.
\bibitem{dar}C.Darwin,Proc.R.Soc.London{\bf 249},180,1959.
\bibitem{bekenstein}J.D.Bekenstein and R.H.Sanders,Astrophys.J{\bf 429},480,1994.
\bibitem{fri}S.Frittelly,T.P.Kling and E.T.Newman,Phys.Rev.D{\bf 61},064021,2000.
\bibitem{ks}K.S.Virbhadra and George F.R.Ellis,Phys.Rev.D{\bf 62},084003,2000.
\bibitem{majum}N.Mukherjee and A.S.Majumdar,G.R.G{\bf 39},583,2007.
\bibitem{rom}E.F.Eiroa,G.E.Romero and D.F.Torres,Phys.Rev.D{\bf 66},024010,2002.
\bibitem{bh}A.Bhadra,Phys.Rev.D{\bf 67},103009,2003.
\bibitem{john}John Southworth et al,Mon.Not.Roy.Astron.Soc.396:1023-1031,2009.
\bibitem{morganson}Eric Morganson, Phil Marshall, Tommaso Treu, Tim Schrabback, Roger D. Blandford, arXiv:0908.0602 [astro-ph.CO]   
\bibitem{song}Song-bai Chen, Ji-liang Jing,Phys.Rev.D{\bf 80},024036,2009. 
\bibitem{koop}L.V.E. Koopmans et al. arXiv:0902.3186 [astro-ph.CO]
\bibitem{faure} C. Faure, J.-P. Kneib, S. Hilbert, R. Massey, G. Covone, A. Finoguenov, A. Leauthaud, 
J.E. Taylor, S. Pires, N. Scoville,  Astrophys.J.{\bf 695},1233,2009. 
\bibitem{bis} G.S. Bisnovatyi-Kogan, O.Yu. Tsupko,Astrophysics {\bf 51},99,2008.
\bibitem{gao}Gao and Zhang,Phys.Rev.D{\bf 70},124019,2004.
\bibitem{sami}Sami,Toporensky,Tretjakov and Tsujikawa,Phy.Lett.B{\bf 619},193,2005.
\bibitem{gas}M.Gasperini,hep-th/0310293.
\bibitem{vir2}C.-M.Claudel,K.S.Virbhadra and G.F.R.Ellis,J.M.P{\bf 42},818,2001.
\bibitem{vir3}K.S.Virbhadra,D.Narasimha and S.M.Chitre,Astron.Astrophys{\bf 337},1-8,1998.
\bibitem{janis}A.I.Janis,E.T.Newman and J.Winicour,Phys.Rev.Lett{\bf 20},878,1968.
\bibitem{boz}V.Bozza,Phys.Rev.D{\bf 66},103001,2002.
\bibitem{vir4}K.S.Virbhadra,Phys.Rev.D{\bf 79},083004,2009.
\bibitem{vir5}K.S.Virbhadra and C.R.Keeton,Phys.Rev.D{\bf 77},124014,2008.
\bibitem{randall}L.Randall and R.Sundrum,Phys.Rev.Lett{\bf 83},4690,1999.
\bibitem{her}S.W.Hawking,T.Hertog and H.S.Reall,Phys.Rev.D{\bf 62},043501,2000.
\bibitem{flachi}Antonino Flachi and Takahiro Tanaka,Phys.Rev.D{\bf 78},064011,2008.
\bibitem{sddmssg} S. Das, D. Maity and S. Sengupta, JHEP {\bf 05}, 042 (2008). 
\end{thebibliography}
\end{document}